# The Use of Radio Observations to Probe Ambient Gas Densities

By Greg F. Wellman & Ruth A. Daly

Department of Physics, Princeton University, Princeton, NJ 08544

The radio properties of powerful extended radio sources may be used to estimate the ambient gas density in the vicinity of radio lobes. A sample of 27 radio lobes from 14 radio galaxies and of 14 radio lobes from 8 radio loud quasars was constructed using sources from the published literature with sufficient radio information to allow an estimate of the ambient gas density. The ambient gas density as a function of separation of the lobe from the center of the parent galaxy indicates a composite density profile, where Cygnus A plays the key role of determining the normalization for the ambient gas density. The composite density profile of the galaxies and quasars studied here is similar to the density profile of gas in low-redshift clusters of galaxies, which confirms the result obtained by Daly [1] using a somewhat smaller sample of radio sources. The data presented here allow an estimate of the core gas density, core radius, and slope of the density profile assuming that the gas density can be fit by a King model.

The data suggest that the core gas density decreases as the redshift of the source increases, and is consistent with either a roughly constant core radius, or a core radius that increases with increasing source redshift. Thus, our results are completely consistent with observations indicating negative evolution of the cluster X-ray luminosity function, and with those indicating that the X-ray luminosities of rich clusters at modest redshift (about 0.5) are significantly lower than those of optically similar zero-redshift clusters.

The negative evolution of the core gas density estimated here is consistent with either a constant total gas mass in which the central density increases and the core radius decreases with time, or a central density that increases with time while the core radius remain roughly constant, in which case the gas mass would increase with time.

Ambient gas densities estimated using the radio properties of powerful extended radio sources provide an useful complement to ambient gas densities estimated from X-ray observations, Faraday rotation measure studies, and measurements of the Sunyaev-Zel'dovich effect. As described here and by [1,2,3,4], radio observations allow an estimate of the ambient gas density as a single parameter for each source; this density could be combined with X-ray observations, Faraday rotation measures, and Sunyaev-Zel'dovich measurements to estimate the size of the X-ray emitting region, the line integral of the magnetic field through the gas about the radio source, and to constrain the slope of the density profile. In addition, since powerful extended radio sources are observed out to very high redshift, this method can be used to probe ambient gas densities of sources that would be very difficult to detect as X-ray sources.

## 1. Introduction

Powerful extended radio sources serve as markers that indicate sites of activity throughout the universe. These sources pinpoint locations in the distant universe that can be studied to learn about the structure of the universe when it was much younger than it is today and to study evolution of structure in the universe, and thus provide one of the very few tools available at present to locate and study the properties of the distant universe.

One of the many interesting aspects of powerful extended radio sources that can be studied is the ambient gas density in the vicinity of the radio lobes of the source [1,3]. Very powerful FRII radio sources are thought to be propagating supersonically relative to the sound speed of the ambient gas as discussed, for example, by [5]. This implies that the strong shock jump conditions apply across the lobe-medium interface. Thus, the radio properties of a radio source may be used to estimate the ambient gas density [1,2,3]. The ambient gas densities indicated using this technique suggest that very powerful FRII sources are in a cluster-like gaseous environment [1,3,4] and the evolution of the gaseous environment of the sources with redshift may indicate the evolution of the intracluster medium (ICM). It is interesting to note that at a redshift of about 2, these sources have a comoving number density comparable to that of rich clusters of galaxies at the present epoch if the short lifetime of the radio source relative to the Hubble time is taken into account, as discussed in §9 of [6].

Since clusters of galaxies are the most massive gravitationally bound systems in the universe, the evolution of these systems with redshift provides a very powerful way to distinguish between various models of structure formation and evolution as discussed, for example, by [7]. Indeed, the strong negative evolution of the cluster X-ray luminosity function with redshift has been taken to indicate strong evolution of the mass function of clusters of galaxies [8,9], as has been the presence of X-ray substructure of low-redshift clusters [10]. However, it is possible that X-ray evolution and the presence of substructure is not due to the gravitational growth of the cluster, but rather is due to effects internal to the cluster such as the relaxation of the gas in the dark matter potential of the cluster, addition of gas to the ICM from cluster members, the cooling and settling of gas, etc.

It appears that the radio properties of powerful extended radio sources may be used to study the gaseous environments of the sources and the evolution of these environments with redshift. This can be combined with optical studies of the environments of the sources to compare the evolution of the gaseous environments with the evolution of the galactic environment of the sources. This may give some indication of whether the negative evolution of the ICM with redshift is due to the gravitational growth of the cluster or due to processes internal to the cluster. It may also give an indication of the formation history of the cluster relative to that of the ICM. The very distant goal of this study is to use the evolution of clusters of galaxies as a discriminator between models of structure formation and evolution.

Values of the Hubble constant of 100 km s$^{-1}$ Mpc$^{-1}$ and deceleration parameter $q_o = 0$ are assumed throughout; results obtained assuming $q_o = 0.5$ are mentioned in §3.

## 2. Ambient Gas Densities

The ambient gas density in the vicinity of the radio lobes of powerful extended radio sources can be estimated by applying the strong shock jump conditions across the lobe-medium interface [1,2,3]. The strong shock jump conditions imply that the lobe is ram pressure confined, so that the lobe pressure $P_L$ is proportional to the ambient gas density $n_a$ and the lobe propagation velocity $v_L$:

$$P_L \propto n_a\, v_L^2 \, . \tag{1}$$

The lobe pressure can be estimated using the minimum energy magnetic field of the radio source. This is a good approximation if the relativistic electrons and magnetic field in the lobes of the sources are in rough pressure equilibrium, or if the deviations from equilibrium are similar from source to source. For example, detailed observations of Cygnus A [11] and of 3C295 [12] suggest that the lobe magnetic fields in each of these sources are about a factor of three less than the minimum energy magnetic field. The data presented here suggest that it is likely that all lobes deviate from minimum energy conditions in the same way, as discussed in detail by [4]. If the minimum energy magnetic field of the lobes is a good approximation to the true field strength, or if deviations from minimum energy are similar from source to source, the lobe pressure is proportional to the square of the minimum energy magnetic field so that [1,2,3,4]

$$n_a \propto \left(\frac{B_{min}}{v_L}\right)^2 \, . \tag{2}$$

The lobe propagation velocity can be estimated from the change of the radio spectral index $\alpha$ across the radio bridge of the source, as discussed in detail by [13,14,15,16], and by [11] for the case of Cygnus A.

Eq. (2) was used by [2] to eliminate the ambient gas density as a variable in the use of powerful extended radio sources as a cosmological tool (see the paper by Guerra and Daly in this volume for an update), and by [1,3] to study the ambient gas density in the vicinity of radio lobes of powerful extended FRII sources. The sources included in these studies were all very large sources and thus could be used to probe the ambient gas density at relatively large distances from the center of the parent galaxy.

The present study was undertaken to obtain ambient gas densities over a broader range of distances from the galaxy core, and to tailor the data analysis to obtain exactly the parameters needed for eq. (2) rather than using related parameters as was done by [1,2].

In order to be able to use radio data to estimate the lobe propagation velocity it is necessary to have fairly high-resolution data extending to relatively low surface brightness levels for at least 2 radio frequencies. This allows an estimate of the change of the radio spectral index across the radio bridge. The change in the radio spectral index is interpreted as the result of synchrotron and inverse Compton aging of the relativistic electrons, and indicates the time that has elapsed since the source was smaller. The time interval that has passed across a given region of the radio source is used to estimate the rate of growth of the source, known as the lobe propagation velocity.

The lobe propagation velocity and minimum energy magnetic field has been estimated for many sources, both galaxies and quasars, from the samples of Leahy, Muxlow, and Stephens [15] and Liu, Pooley, and Riley, [16], as described by Wellman & Daly [4]. The minimum energy magnetic field was estimated both at 10 and 25 kpc from the hotspot in the direction of the host galaxy. The 10 kpc results will be discussed here.

The ambient gas density obtained using eq. (2) is plotted as a function of the separation between the estimated center of the host galaxy and the hotspot in figure 1a. The normalization of eq. (2) is set using Cygnus A with the following parameters: the ambient gas density is about $1.6 \times 10^{-2}$ cm$^{-3}$ [18] at the position of the radio lobes, the minimum energy magnetic field is $9 \times 10^{-5}$ gauss [4,11], and the lobe propagation velocity derived with that field is $0.08c$ [4,11].

The composite density profile shown in figure 1a is very similar is shape to that of the ICM in low redshift clusters of galaxies, and its normalization is similar to that of low-redshift clusters by definition it is set using Cygnus A. The composite density profile has been fit with a density profile of the form:

$$n_a(r) = n_c \left(1 + \left(\frac{r}{r_c}\right)^2\right)^{-3\beta/2}, \qquad (3)$$

where $n_c$ is the core density, $r_c$ is the core radius, and $\beta$ indicates the slope of the density profile (e.g. Sarazin [17] eq. [5.63]). The best fit parameters and the reduced $\chi^2$ of the fit are listed on the figure. The three dimensional error ellipsoid has a complicated shape [4]. A two dimensional error ellipsoid is obtained by fixing $\beta$ to the best fit value of 0.9, and indicates uncertainties on the remaining two parameters of $n_c = (11^{+6}_{-4}) \times 10^{-3}$ cm$^{-3}$ and $r_c = 90^{+60}_{-20}$ kpc.

The value of the reduced $\chi^2$ for the fit shown in figure 1a is rather large. If the sample is divided into high and low-redshift subsamples it is clear that the core density of the $z > 1$ sources is lower than the $z < 1$ sources, while the redshift evolution of the core radius is less clear (see [4] for more details). This motivates a fit to the composite density profile that explicitly includes evolution with redshift.

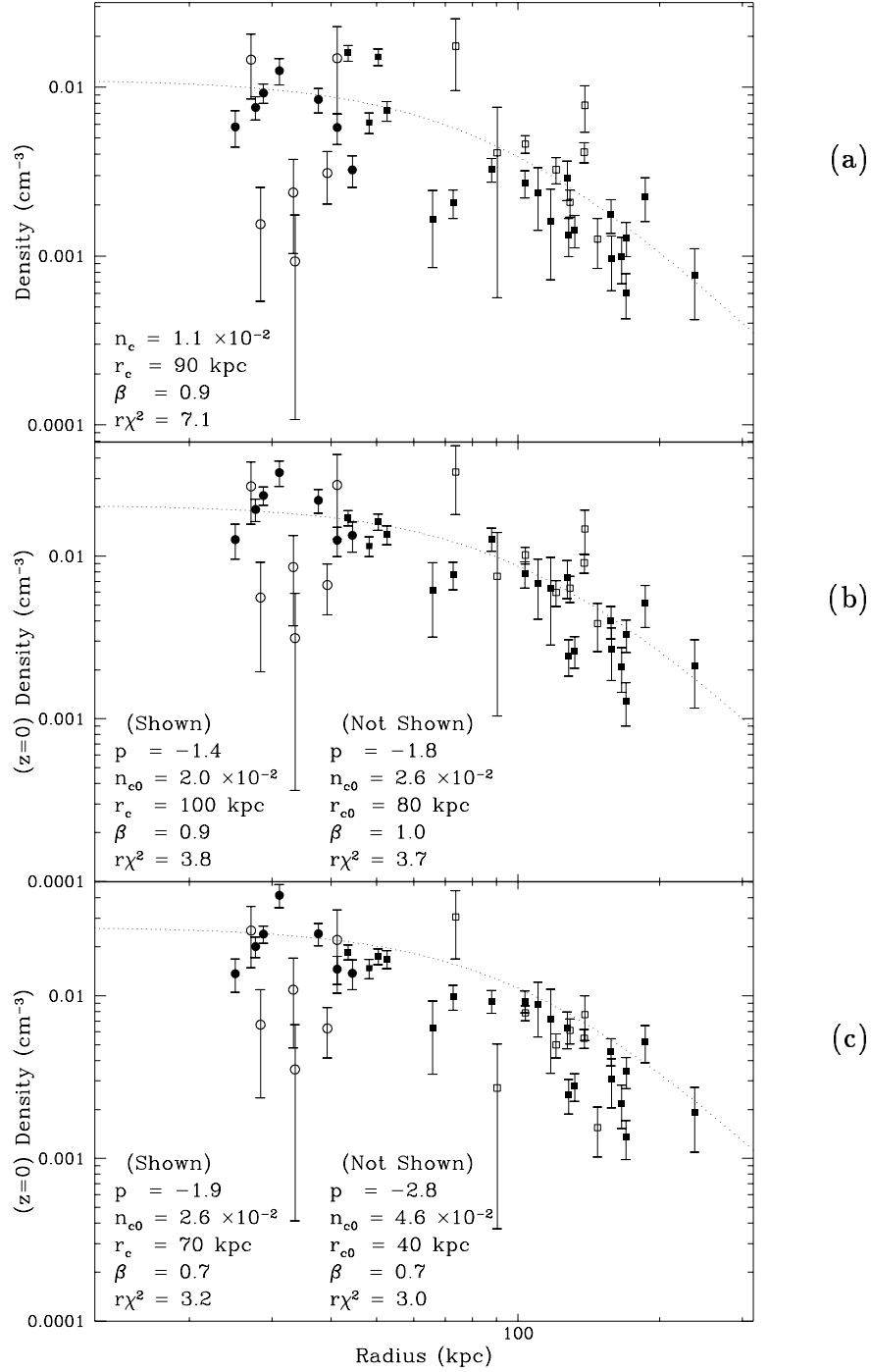

FIGURE 1. (a) Results for 10 kpc data with no evolution and $B = B_{min}$; (b) Results for 10 kpc data with evolution and $B = B_{min}$; (c) Results for 10 kpc data with evolution and $B = B_{min}/3$.

### 2.1. *Evolution of Ambient Gas Densities with Redshift*

Two forms for the evolution of the ambient gas density with redshift are presented here; both assume that the density profile is well described by a King model, and both require that 4 parameters be constrained by the data. The first is identical to eq. (3) with $n_c$ replaced by $n_{c0}(1+z)^p$, and second has this alteration to eq. (3) and $r_c$ is replaced by $r_{c0}(1+z)^{-p/3}$. The first form assumes that the core radius is independent of redshift while the core density changes with redshift, and the second form assumes that the core gas mass is constant while the core radius and density change. Another possibility is that the core radius, core density, and core gas mass change with redshift, but this would require a 5 parameter fit to the data and is not considered here.

Fits allowing for the evolution of $n(r)$ with redshift are shown in figures 1b and 1c. The fits shown are for the first type of evolution considered and described above, and the parameters for the second type of fit are listed on the figure under the caption "not shown." Figures 1b and 1c have been computed assuming $B = B_{min}$ and $B = B_{min}/3$ respectively.

Negative redshift evolution is indicated by all 4 fits, though the magnitude of the evolution varies somewhat. The values of the reduced $\chi^2$s of the fits are significantly lower than obtained assuming no evolution of the density with redshift (compare figure 1a with figures 1b and 1c).

Similar results obtain when a value for the deceleration parameter $q_o = 0.5$ is assumed [4], though the negative evolution with redshift is slightly less significant than the result presented here for $q_o = 0$.

## 3. Discussion

These results and those obtained by [1] suggest that very powerful extended radio galaxies and radio loud quasars lie in similar gaseous environments, which resemble those observed in clusters of galaxies, and suggest that these very powerful extended radio sources lie in the cores of clusters of galaxies. The relatively small core radii indicated by the composite density profile hints that these radio sources may be associated with cD galaxies, since the core radii of the gas in clusters containing cD's is comparable to the core radii obtained here, and is a factor of 2 smaller than gas in clusters lacking a cD galaxy [19].

This study and that of [1] clearly show that the extended radio sources selected as described by [1,2,4] provide a very powerful tool to study the gaseous environments of the sources, and the evolution of these gaseous environments. The results presented here and by [4] suggest negative evolution of the core gas density with redshift. This is consistent with the negative evolution of the cluster X-ray luminosity function [20,21], the low X-ray luminosities of optically rich clusters of galaxies at modest redshift [22], and the evolution of the gaseous environments of less powerful FRII sources with redshift [23,24], all of which indicate that the density of the ICM in clusters of galaxies decreases with increasing redshift.

The evolution of clusters of galaxies with redshift provides a sensitive diagnostic of models of structure formation and evolution. However, in order to be able to use clusters for this purpose, it is necessary to be able to distinguish between evolution associated with the gravitational growth of the cluster and that associated with processes internal to the cluster. Thus, the distant goal of this study is to constrain the evolution of the masses of galaxy clusters and that of the gas that comprises the ICM separately. This, in turn, would have far-reaching implications for models of structure formation and evolution.

The evolution of the cluster gas and that of the cluster gravitational potential could be distinguished by combining the results presented here with studies of the density of galaxies in the vicinity of the radio source, such as those presently underway by Dickinson, Spinrad, and collaborators, for example. Evolution of the cluster mass and the cluster gas could be further constrained using X-ray data, Faraday rotation measure studies, and measurements of the Sunyaev-Zel'dovich effect in distant clusters of galaxies. X-rays from the intracluster medium can be used to indicate the product of the total gas mass and the gas density; Faraday rotation measure studies indicate the product of the gas density, the magnetic field strength, and the extent of the gas; and SZ measurements indicate the product of the gas pressure and the extent of the gas. Combinations of different measures of the state of the gas could provide interesting constraints on the evolution of the ICM in clusters of galaxies.

It is a pleasure thank Roger Blandford, Ed Groth, Eddie Guerra, Tom Herbig, Rick Perley, and Lin Wan for helpful discussions. This work was supported in part by the US National Science Foundation, and NSERC of Canada.